# Suppression of the emittance growth induced by coherent synchrotron radiation in triple-bend achromats

Huang Xi-Yang(黄玺洋), Jiao Yi(焦毅), Xu Gang(徐刚), Cui Xiao-Hao(崔小昊)

Institute of High Energy Physics, Chinese Academy of Sciences, Beijing 100049, China

**Abstract:** The coherent synchrotron radiation (CSR) effect in a bending path plays an important role in transverse emittance dilution in high-brightness light sources and linear colliders, where the electron beams are of short bunch length and high peak current. Suppression of the emittance growth induced by CSR is critical to preserve the beam quality and help improve the machine performance. It has been shown that the CSR effect in a double-bend achromat (DBA) can be analyzed with the two-dimensional point-kick analysis method. In this paper, this method is applied to analyze the CSR effect in a triple-bend achromat (TBA) with symmetric layout, which is commonly used in the optics designs of energy recovery linacs (ERLs). A condition of cancelling the CSR linear effect in such a TBA is obtained, and is verified through numerical simulations. It is demonstrated that emittance preservation can be achieved with this condition, and to a large extent, has a high tolerance to the fluctuation of the initial transverse phase space distribution of the beam.

## I. Introduction

Electron beams with high peak current (on the kilo-ampere scale) and short bunch length (on the sub-picosecond scale) are desired in high-brightness light sources [1-4] and linear colliders [5]. In these machines, when an electron bunch passes through a bending magnet, the emission of the coherent synchrotron radiation (CSR) induces energy modulation along the bunch and dilutes transverse emittance, leading to degradation of the beam quality. This has strongly motivated theoretical [6-9] and numerical analyses [10-18] on CSR over the past few decades.

One important topic among these studies is suppressing the CSR-induced emittance growth in achromats. Several theoretical methods have been proposed, such as the R-matrix analysis [10-11] and the Courant-Snyder (C-S) formalism analysis [14], to evaluate the CSR effect. In the method of R-matrix analysis, 5-by-5 transfer matrices are applied to calculate particle coordinate deviations and emittance growth due to CSR in a linear regime. However, it is difficult to obtain a generic scheme to suppress the CSR-induced emittance growth with this method since the calculated CSR wake dispersion (i.e., the CSR-induced orbit deviation at the exit of the achromat) is different for different concrete lattice designs. On the other hand, in the C-S formalism analysis, a single kick approximation of the CSR effect is used, by assuming the CSR-induced orbit deviation after going through a bending magnet to be $\Delta X_{csr} = (\Delta x, \Delta x') = (D_x, D_x')\delta(csr)$, where $D_x$ and $D_x'$ are the dispersion function and its derivative with respect to $s$ at the center of the dipole, and $\delta(csr)$ is the CSR-induced energy deviation in a bending magnet. This rough approximation makes it feasible to analyze the emittance growth, while resulting in somewhat different results from the more rigorous R-matrix analysis.

To formulate the CSR-induced emittance growth in both a rigorous and an explicit way, a novel method, named 2D point-kick analysis [16-17], was recently proposed. This method adopts a 2D point-kick model of the CSR effect in a bending magnet [see Eq. (2) below], with which one can analyze the CSR-induced emittance growth in only the $(x, x')$ 2D planes (similar to the C-S formalism analysis), and obtain the same result as from the R-matrix analysis. Moreover, in this analysis, the transfer matrix of the beam line between adjacent dipoles is treated as a whole, instead of considering the elements one-by-one (as in the R-matrix analysis). As a result, the CSR kick can be expressed in an explicit way, and the solution for zero CSR-kick provides a rather general requirement on the optics design of the achromat. This method will be briefly reviewed as follows.

In the "steady-state" ($\rho/\gamma^3 \ll \sigma_z \ll \rho\theta/2\gamma^2 + \rho\theta^3/24$) approximation for a Gaussian line-charge distribution beam, the CSR-induced rms relative energy spread depends linearly on both $L_b$ and $\rho^{2/3}$ [6, 18]

$$\Delta E_{rms} = 0.2459 \frac{eQ\mu_0 c_0^2 L_b}{4\pi\rho^{2/3}\sigma_z^{4/3}}, \tag{1}$$

where $\gamma$, $e$, $Q$, $\rho$, $\sigma_z$, $L_b$, $\mu_0$, $c_0$ represent the relativistic Lorentz factor, the charge of a single particle, the bunch charge, the bending radius of the orbit, the rms bunch length, the bending path, the permeability of vacuum, and the speed of light, respectively. It has been verified through ELEGANT simulations [16] that this relation applies well to the cases with $\theta$ ranging from 1 to 12 degrees and $\rho$ ranging from 1 to 150 m.

Therefore the CSR effect in a dipole was linearized by assuming $\delta(csr) = kL_b/\rho^{2/3}$, where $k$ depends only on the bunch charge $Q$ and the bunch length $\sigma_z$, and is in unit of $m^{1/3}$. In addition, it was shown that the CSR-induced coordinate deviations after a passage through a dipole can be equivalently formulated with a point-kick at the center of the dipole (see Fig. 1), which is of the form [16]

$$X_k = \begin{pmatrix} \rho^{4/3} k[\theta\cos(\theta/2) - 2\sin(\theta/2)] \\ \sin(\theta/2)(2\delta + \rho^{1/3}k\theta) \end{pmatrix}, \tag{2}$$

where $\delta = \delta_0 + \delta_{csr}$, is the particle energy deviation at the entrance of the dipole, with $\delta_0$ being the initial particle energy deviation and $\delta_{csr}$ being that caused by CSR in the upstream path.

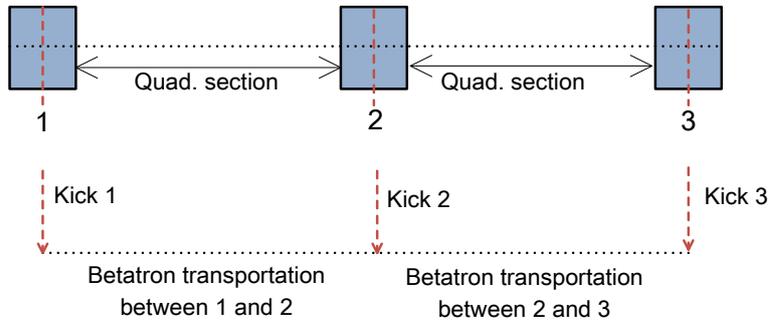

Fig. 1 Schematic layout of a symmetric TBA and the corresponding physical model of the CSR effect in a TBA with three point-kicks.

Using this point-kick model and treating the transportation of a particle between adjacent kicks as a whole, the CSR-induced coordinate deviations in an n-dipole achromat can be evaluated, by analyzing the horizontal betatron motion with *n*-point kicks. With this method, the condition of cancelling the net CSR kick in a double-bend achromat (DBA) with two identical dipoles has been obtained [16],

$$M_{c2c} \cong \begin{pmatrix} -1 & 0 \\ 12/L_b & -1 \end{pmatrix}, \tag{3}$$

where $M_{c2c}$ is 2-by-2 transfer matrix for the beam line between the centers of the two dipoles of a DBA, and $L_b$ is the length of the dipole.

In this paper, the 2D point-kick analysis method is expanded to study the CSR effect in a triplet-bend achromat (TBA) with symmetric layout, and the linear CSR effect cancellation condition is obtained. Details are presented in Sec. II, then in Sec. III, the condition found is verified with ELEGANT simulations. It turns out that the proposed condition is quite robust against the mismatch of beam optics. Note that TBA is the minimum configuration required to realize an isochronous cell (with zero $R_{56}$) and is usually adopted in the design of the recirculation loop of energy recovery linacs (ERLs, see, e.g., [19, 20]). In most cases, the bunch length is rarely changed in these TBAs. For example, in both HC (high current) and LE (low emittance) modes of the compact energy recovery linac (ERL) in Japan, the variations of the bunch length due to 0.1% RF voltage and phase error are below 1% in the first TBA [21]. Thus, this method is reliable (which based on the assumption that the bunch length is constant in a dipole) and the presented results will be important to preservation of the beam quality in those high-brightness ERLs.

**II. Generic linear CSR-Cancellation Condition for a symmetric TBA**

In this section, we will present the derivation of the linear CSR-cancellation condition for a TBA with symmetric layout (i.e., the distribution and strength of the elements are all symmetric with respect to the center of the TBA). Such a TBA is chosen based on the following considerations: a symmetric TBA is always adopted in practical designs of the recirculation loop of the ERLs; and with symmetric layout, only a few decision variables need to be determined, which makes it feasible to find explicit achromatic and linear CSR-cancellation condition. As sketched in Fig. 1, the bending angles of the first and the third dipoles are the same (denoted by $\theta_1$), while that of the second dipole can be arbitrary (denoted by $\theta_2$); the bending radii of these dipoles are the same (denoted by $\rho$). According to the 2D point-kick analysis, CSR kicks occur at the centers of the three dipoles (denoted by 1, 2, 3, in Fig. 1), and between the adjacent kicks only one 2-by-2 transfer matrix of the horizontal betatron motion is considered. Without loss of generality, the transfer matrix between point 1 and 2 is expressed as

$$M_{12} = \begin{pmatrix} m_{11} & m_{12} \\ m_{21} & m_{22} \end{pmatrix}, \tag{4}$$

where the symplectic condition should be satisfied, i.e., $\det M_{12} = 1$. Due to the symmetric layout, the matrices for the two half sections are related to each other. The matrix between point 2 and 3 (denoted by $M_{23}$) is given by [22]

$$M_{23} = \begin{pmatrix} m_{22} & m_{12} \\ m_{21} & m_{11} \end{pmatrix}. \tag{5}$$

With these matrices and the CSR point-kicks, one can evaluate the particle coordinate deviations in relation to $\delta_0$ and $k$, respectively, and the corresponding emittance growth after passage through the TBA. For simplicity, it is assumed that the initial particle coordinates relative to the reference trajectory are $X_0 = (x_0, x_0')^T = (0, 0)^T$ and the energy deviation is $\delta = \delta_0$ (this assumption is reasonable since the betatron motion of the particle will not lead to additional emittance growth).

The coordinates remain zero until the particle experiences the CSR-kick at point 1, where the particle coordinates are given by

$$X_1 = X_0 + X_{k1} = \begin{pmatrix} \rho^{4/3} k[\theta_1 \cos(\theta_1/2) - 2\sin(\theta_1/2)] \\ \sin(\theta_1/2)(2\delta_0 + \rho^{1/3} k\theta_1) \end{pmatrix} = \begin{pmatrix} -\rho^{4/3} k r_1 \\ S_1(2\delta_0 + \rho^{1/3} k\theta_1) \end{pmatrix}. \tag{6}$$

where $r_1 = 2\sin(\theta_1/2) - \theta_1 \cos(\theta_1/2)$, and $S_1 = \sin(\theta_1/2)$. After passing through the section between point 1 and 2, the particle experiences the second kick,

$$\begin{aligned} X_2 &= \begin{pmatrix} x_2 \\ x'_2 \end{pmatrix} = M_{12} X_1 + X_{k2} \\ &= \begin{pmatrix} k\rho^{1/3}[-(r_2 + r_1 m_{11})\rho + m_{12}\theta_1 S_1] + 2m_{12} S_1 \delta_0 \\ k\rho^{1/3}[-r_1 m_{21}\rho + m_{22}\theta_1 S_1 + (2\theta_1 + \theta_2) S_2] + (2m_{22} S_1 + 2S_2)\delta_0 \end{pmatrix}, \end{aligned} \tag{7}$$

where $r_2 = 2\sin(\theta_2/2) - \theta_2 \cos(\theta_2/2)$, and $S_2 = \sin(\theta_2/2)$. Similarly, the particle coordinate deviations at point 3 (after the third kick) are

$$\begin{aligned} X_3 &= \begin{pmatrix} x_3 \\ x'_3 \end{pmatrix} = M_{23} X_2 + X_{k3} \\ &= \delta_0 \begin{pmatrix} 2m_{12}(2m_{22} S_1 + S_2) \\ 2m_{11}(2m_{22} S_1 + S_2) \end{pmatrix} + k\rho^{1/3} \begin{pmatrix} [-m_{22}(r_2 + 2r_1 m_{11})\rho + 2m_{22} m_{12}\theta_1 S_1 + m_{12}(2\theta_1 + \theta_2) S_2] \\ [-(r_2 + 2r_1 m_{11})\rho m_{21} + 2(\theta_1 + m_{11} m_{22}\theta_1 + \theta_2) S_1 + m_{11}(2\theta_1 + \theta_2) S_2] \end{pmatrix}. \end{aligned} \tag{8}$$

Note that in $X_{k2}$ and $X_{k3}$ [in Eqs. (7) and (8)], the energy spread $\delta$ grows to $k\rho^{1/3}\theta_1 + \delta_0$ and $k\rho^{1/3}(\theta_1 + \theta_2) + \delta_0$, respectively.

One can see that the overall particle coordinate deviations include two parts: the momentum dispersive terms related to the initial energy deviation $\delta_0$ [the first term on the right-hand side of Eq. (8)] and the CSR-dispersive terms related to $k$ [the second term on the right-hand side of Eq. (8)]. The above equation can therefore be written as

$$\begin{aligned} x_3 &= \Delta x_3(\delta_0) + \Delta x_3(csr), \\ x'_3 &= \Delta x'_3(\delta_0) + \Delta x'_3(csr). \end{aligned} \tag{9}$$

The conventional achromatic condition can be obtained when $\Delta x_3(\delta_0)$ and $\Delta x'_3(\delta_0)$ are equal to zero,

$$m_{22} = -\frac{S_2}{2S_1}. \tag{10}$$

This indicates that in a TBA with symmetric layout, the achromatic condition only depends on the bending angles of the dipoles.

Substituting the achromatic condition and the symplectic condition into Eq. (9), the CSR induced coordinate deviations in a TBA with symmetric layout can generally be expressed as

$$x_3 = \Delta x_3(csr) = \frac{1}{2}k\rho^{1/3}[2m_{12}(\theta_1 + \theta_2) + (r_2 + 2r_1 m_{11})\rho / S_1]S_2,$$
$$x'_3 = \Delta x'_3(csr) = \frac{1}{2}k\rho^{1/3}[2m_{12}(\theta_1 + \theta_2) + (r_2 + 2r_1 m_{11})\rho / S_1]\frac{2S_1 + m_{11}S_2}{m_{12}}. \tag{11}$$

In storage rings where the typical bunch length of the electron beam is usually on the scale of 1 cm, the CSR effect is too small to have an impact on the beams. However, in linac-driven ERLs with very short beams, the CSR effect cannot be ignored, and if not well suppressed, will cause evident growth in the geometric emittance.

The final geometric emittance (the geometric emittance at the end of the TBA) in presence of the CSR effect can be estimated by

$$\varepsilon^2 = (\varepsilon_0 \beta_x + \Delta x_{3,rms}^2)(\varepsilon_0 \gamma_x + \Delta x'^{2}_{3,rms}) - (\varepsilon_0 \alpha_x - \Delta x_{3,rms}\Delta x'_{3,rms})^2 = \varepsilon_0^2 + \varepsilon_0 \cdot d\varepsilon,$$
$$d\varepsilon = \gamma_3 \Delta x_{3,rms}^2 + 2\alpha_3 \Delta x_{3,rms}\Delta x'_{3,rms} + \beta_3 \Delta x'^{2}_{3,rms}, \tag{12}$$

where $\varepsilon_0$ is the unperturbed geometric emittance and $\alpha_3$, $\beta_3$, $\gamma_3$ are the C-S parameters at the center of the third dipole of the TBA. In most cases $d\varepsilon \ll \varepsilon_0$, therefore the growth in unnormalized and normalized emittance due to CSR can be estimated by

$$\Delta\varepsilon = \varepsilon - \varepsilon_0 \approx \frac{1}{2}d\varepsilon,$$
$$\Delta\varepsilon_n = \varepsilon_n - \varepsilon_{n0} = \gamma\beta(\varepsilon - \varepsilon_0) \approx \frac{1}{2}\gamma\beta d\varepsilon, \tag{13}$$

where $\beta$ is the particle velocity relative to the speed of light, $\gamma$ is the relativistic Lorentz factor, and the subscript $n$ represents the normalized emittance. To achieve a zero emittance growth [see Eqs. (11) and (12)], it is required that $[\Delta x_3(csr), \Delta x'_3(csr)]^T = (0, 0)^T$, from which the linear CSR-cancellation condition can be obtained,

$$m_{11} + \frac{(\theta_1 + \theta_2)S_1}{r_1 \rho}m_{12} = -\frac{r_2}{2r_1}. \tag{14}$$

In the particular case with $\theta_1 \ll 1$ and $\theta_2 \ll 1$, Eq. (14) can be simplified as

$$m_{11} + \frac{6(\theta_1 + \theta_2)}{\theta_1^2 \rho}m_{12} \cong -\frac{\theta_2^3}{2\theta_1^3}, \tag{15}$$

where only the first significant terms with respect to $\theta_1$ and $\theta_2$ are kept.

Note that Eq. (14) [or Eq. (15)] only imposes a general requirement of the transfer matrix of the betatron transportation section, which suggests a generic and easily-applied way to suppress the CSR-induced emittance growth in a symmetric TBA, i.e., varying the quadrupole strengths (and the position, if necessary) to give a transfer matrix $M_{12}$ in the form of

$$M_{12} = \begin{pmatrix} -\dfrac{r_2\rho + 2m_{12}(\theta_1+\theta_2)S_1}{2r_1\rho} & m_{12} \\ \dfrac{1}{m_{12}}(\dfrac{r_2 S_2}{4r_1 S_1} + \dfrac{m_{12}(\theta_1+\theta_2)S_2}{2r_1\rho} - 1) & -\dfrac{S_2}{2S_1} \end{pmatrix}, \quad (16)$$

where only one term $m_{12}$ is variable. In addition, this condition is tenable when the TBA degenerates to a DBA by setting $\theta_2 = 0$. The transfer matrix between the centers of the first and the third dipoles can be evaluated, giving the same form as in Eq. (3),

$$M_{c2c} = M_{23} M_{12} = \begin{pmatrix} 0 & m_{12} \\ -1/m_{12} & -m_{12}\theta_1 S_1/r_1\rho \end{pmatrix} \begin{pmatrix} -m_{12}\theta_1 S_1/r_1\rho & m_{12} \\ -1/m_{12} & 0 \end{pmatrix}$$
$$= \begin{pmatrix} -1 & 0 \\ 2\theta_1 S_1/r_1\rho & -1 \end{pmatrix} \cong \begin{pmatrix} -1 & 0 \\ 12/L_b & -1 \end{pmatrix}. \quad (17)$$

where only the first significant terms with respect to $\theta_1$ are kept.

### III. Theoretical Verifications and Numerical Simulations

In this section, we will verify the proposed linear CSR-cancellation condition with ELEGANT simulations. In these simulations, we consider a TBA consisting of three identical dipoles with bending radii of 7 m and bending angles of 3°. Under this circumstance, the element $m_{22}$ of the transfer matrix is equal to -0.5 [according to Eq. (10)], and the other elements are controlled by varying the quadrupole strengths. The main parameters of the electron beam for simulations are listed in Table 1.

Table 1. Parameters of the electron beam and the TBA used in the ELEGANT simulations

| Parameter | Value | Units |
|---|---|---|
| Bunch charge | 500 | pC |
| Norm. emittance | 2 | μm.rad |
| Beam energy | 1000 | MeV |
| Energy spread | 0.05 | % |
| Bunch length | 30 | μm |
| Dipole bending radius | 7 | m |
| Dipole bending angle | 3 | degree |

In order to confirm that the condition found can result in minimum emittance growth, we investigate the dependency of the final emittance on $m_{11}$ by fixing $m_{12}$. The simulation results are shown in Fig. 2. In these cases, $m_{11}/m_{11}^*$ is varied from about 0.2 to 2, where $m_{11}^*$ is the solution of Eq. (15) and has the value of 8.058 when $m_{12} = -0.261$ (red points) and 34.127 when $m_{12} = -1.058$ (blue points). In order to make the analytical emittance growth calculation easier, the optics are also designed with symmetric C-S parameters in each case. Through straightforward derivations, the final normalized emittance growth in Eq. (13) can be obtained, and is in the form

$$\Delta\varepsilon_n = 2\gamma\beta k^2 q^2 \rho^{8/3} m_{11}^{*2}(1 - m_{11}/m_{11}^*)^2 / \beta_2. \qquad (18)$$

From Fig. 2 it can be seen that the emittance growth reaches a minimum as $m_{11}$ is on or close to the optimal value, which agrees reasonably well with the analytical prediction. Note that, the minimum $\Delta\varepsilon_n$ is not exactly zero, but on scale of 0.001 μm.rad. This is because the nonlinear effect of the CSR wake is included in the simulations, which becomes dominant as the linear CSR effect is eliminated. However, compared with the rapid increase in $\Delta\varepsilon_n$ as $m_{11}$ deviates away from $m_{11}^*$, the nonlinear effect is rather weak relative to the linear effect of the CSR wake, and therefore with the proposed condition the emittance growth in a TBA can be well controlled.

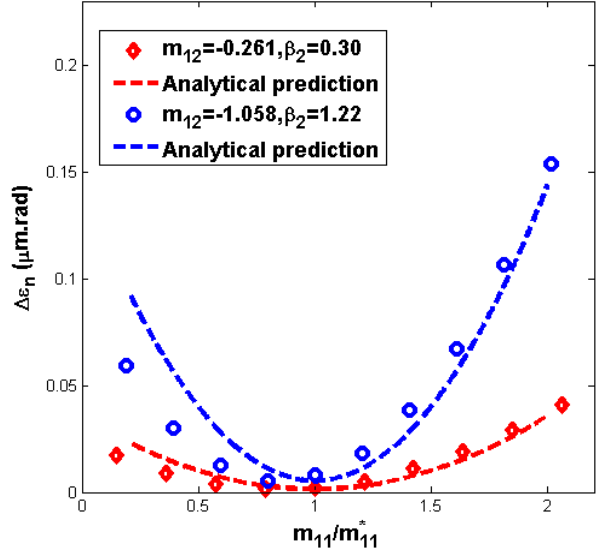

Fig. 2. (color online) Variation of the final normalized emittance through TBA with respect to $m_{11}/m_{11}^*$, for the cases with the term $m_{12}$ being -0.261 (red points) and -1.058 (blue points), respectively. In all cases the achromatic condition is satisfied. The dashed lines are analytical prediction from Eq. (19) and with a shift of the minimum $\Delta\varepsilon_n$.

As mentioned above, the CSR-cancellation condition only imposes a requirement on the transfer matrix of the betatron transportation section. In the following, we will investigate the dependence of emittance suppression on the initial phase space distribution of the beam (or namely the initial C-S parameters). In most cases, symmetric optics are adopted in the design of TBAs in ERLs. For example, the initial C-S parameters of the TBA lattice with symmetric optics (and with $m_{11}$ = -3.297, $m_{12}$ = 0.085 and other parameters as listed in Table 1) are $\beta_0$ = 4.61m and $\alpha_0$ = 30. We vary the initial C-S parameters, and then track the electron beam through the TBA to record the corresponding emittance growth. The results are presented in Fig. 3. One can see that there is a gradual increase in final emittance as the initial C-S parameters deviate from the nominal value. This is probably due to the enhancement of the nonlinear effect of the CSR wake along with the increase of the final C-S parameters [see Fig. 4]. For example, in an extreme case with $\beta_0$ = 4.61m and $\alpha_0$ = 50, the final C-S parameters become too large ($\beta_f$ = 1846 m, $\alpha_f = -11981$) and the final emittance increases more than doubled. However, the relative emittance growth keeps to below 1% when $\beta_0$ is varied by 0.5 m and $\alpha_0$ is varied by 2. This suggests that the proposed CSR-suppression scheme is rather robust against the fluctuation of the initial beam distribution pulse by pulse.

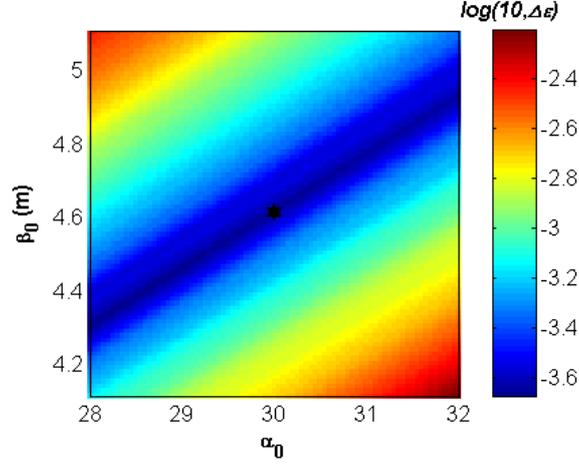

Fig. 3. (color online) Numerical simulations by scanning initial C-S parameters. The CSR-suppression condition [Eqs. (10) & (14)] is ensured for the test TBA, with the element of the transfer matrix $M_{12}$ are (-3.297, 0.085, 7.591, -0.5) for ($m_{11}$, $m_{12}$, $m_{21}$, $m_{22}$), respectively. The black star represents the case with symmetric optics, with $\beta_0 = 4.61$m and $\alpha_0 = 30$.

.

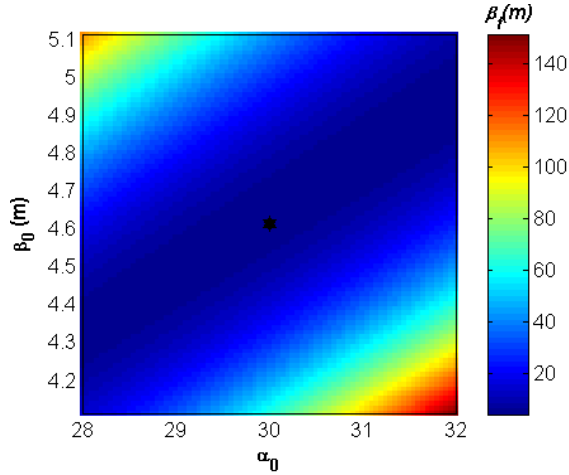

Fig. 4. (color online) The final C-S parameter $\beta_f$ with scanning $\beta_0$ and $\alpha_0$. The black star represents the case with symmetric optics, with $\beta_0 = 4.61$m and $\alpha_0 = 30$.

### IV. Conclusion

In this paper, the 2D CSR-kick analysis method is applied to study the suppression of CSR-induced emittance growth in a TBA cell with symmetric layout, and the explicit achromatic and linear CSR-cancellation condition for the TBA are derived. Since the linear effect of the CSR wake is cancelled, the emittance growth due to CSR can be well controlled with the proposed condition, even with optics mismatch from a nominal symmetric design. Finally, it is worth mentioning that the 2D point-kick analysis assumes the bunch length has little change in the achromat, and thus the results presented in this paper are more appropriate to a transport line with small momentum compactions such as an ERL recirculation loop than to the specified bunch compressors. Further study will be continued and extended in order to explore the case with large variation of the bunch length.